\documentclass[prl,twocolumn,showpacs,preprintnumbers,amsmath,amssymb,citeautoscript]{revtex4-1}

\usepackage{graphicx}
\usepackage{dcolumn}
\usepackage{bm}
\usepackage{color}

\begin{document}

\title{Superconducting gap with sign reversal between hole pockets in heavily hole-doped Ba$_{1-x}$K$_x$Fe$_2$As$_2$}

\author{D.~Watanabe$^1$}
\author{T.~Yamashita$^1$}
\author{Y.~Kawamoto$^1$}
\author{S.~Kurata$^1$}
\author{Y.~Mizukami$^1$}
\author{T.~Ohta$^1$}
\author{S.~Kasahara$^1$}
\author{M.~Yamashita$^{1,2}$}
\author{T.~Saito$^3$}
\author{H.~Fukazawa$^3$}
\author{Y.~Kohori$^3$}
\author{S.~Ishida$^4$}
\author{K.~Kihou$^4$}
\author{C.\,H.~Lee$^4$}
\author{A.~Iyo$^4$}
\author{H.~Eisaki$^4$}
\author{A.\,B.~Vorontsov$^5$}
\author{T.~Shibauchi$^1$}
\author{Y.~Matsuda$^1$}

\affiliation{$^1$Department of Physics, Kyoto University, Kyoto 606-8502, Japan}
\affiliation{$^2$Institute for Solid State Physics, University of Tokyo, Chiba 277-8581, Japan}
\affiliation{$^3$Department of Physics, Chiba University, Chiba 263-8522, Japan}
\affiliation{$^4$National Institute of Advanced Industrial Science and Technology (AIST), Tsukuba, Ibaraki 305-8568, Japan}
\affiliation{$^5$Department of Physics, Montana State University, Bozeman, Montana, 59717, USA}

\date{\today}

\begin{abstract}

To gain insight into the unconventional superconductivity of Fe-pnictides with no electron pockets, we measure the thermal conductivity $\kappa$ and penetration depth $\lambda$ in the heavily hole-doped regime of Ba$_{1-x}$K$_x$Fe$_2$As$_2$.  The residual thermal conductivity $(\kappa/T)_{T \rightarrow 0\,{\rm K}}$ and $T$-dependence of $\lambda$ consistently indicate the fully gapped superconductivity at $x=0.76$ and the (line) nodal superconductivity at higher hole concentrations.  The magnitudes of 
$\frac{\kappa}{T}\cdot T_c|_{T \rightarrow 0\,{\rm K}}$ and $\frac{d\lambda}{d(T/T_c)}$ at low temperatures, 
both of which are determined by the properties of the low-energy excitations, 
exhibit a highly unusual non-monotonic $x$-dependence.  These results indicate a dramatic change of the nodal characteristics
in a narrow doping range, suggesting a doping crossover of the gap function between the $s$-wave states with and without sign reversal between $\Gamma$-centered hole pockets. 

\end{abstract}

\pacs{74.20.Rp,74.25.Dw,74.70.Xa,74.25.fc}


\maketitle

In most iron-based high-$T_c$ superconductors, the interband interaction between the disconnected hole and electron Fermi-surface pockets is believed to be responsible for electron pairing and superconductivity \cite{Paglione10,Stewart11,Hirschfeld11}.   Exceptional cases are heavily hole-doped iron-pnictide Ba$_{1-x}$K$_x$Fe$_2$As$_2$ and heavily electron-doped iron-chalcogenide K$_x$Fe$_{2-y}$Se$_2$, which have only hole pockets and electron pockets, respectively,  but still exhibit superconductivity.  Since the superconducting (SC) gap structure reflects the underlying structure of the pairing interaction,  the gap structure of this class of iron based compounds has been attracted great interest and various kinds of SC gap functions have been put forward theoretically, including $d$, $s+id$ and $s+is$ states \cite{Thomale11,Maiti12,Platt12,Maiti13}.   

In Ba$_{1-x}$K$_x$Fe$_2$As$_2$  the electron pockets disappear at $x\agt 0.6$ \cite{Malaeb12} or $x>0.7$ \cite{Nakayama11} and the Fermi surface consists of three main two dimensional hole pockets at the center of the Brillouin zone and small hole pockets near the zone corner \cite{Terashima10,Malaeb12}.   In contrast to nodeless extended $s$-wave superconductivity in optimally-doped  Ba$_{1-x}$K$_x$Fe$_2$As$_2$ ($T_c=38$\,K, $x\approx0.45$) \cite{Ding08,Hashimoto09,Evt09,Malaeb12},  line nodes appear in fully hole doped  material KFe$_2$As$_2$ \cite{Hashimoto10,Fukazawa09,Dong10,Reid12}, prompting a proposal of a $d$-wave SC gap function based on the thermal conductivity measurements \cite{Reid12}.  However, subsequent laser angle resolved photoemission spectroscopy (ARPES) has revealed a nodal $s$-wave gap with a peculiar structure: a nodeless gap on the inner hole pocket, an unconventional gap with ``octet-line nodes'' on the middle pocket and an almost-zero gap on the outer pocket \cite{Okazaki12}.   This $A_{1g}$ gap symmetry indicates that these line nodes are not symmetry protected but accidental.   

It has been extensively discussed that the interband hopping plays a major role in iron-pnictides, and in compounds with both hole (h) and electron (e) pockets it gives rise to $S_{+-}^{he}$ gap, where the SC order parameter has opposite signs on hole and electron pockets.  Then the question arises, what is the pairing mechanism of superconductivity in pnictides with {\it no} electron pockets? In this respect it is of particular interest to find out whether the sign change of the SC order parameter occurs between the {\it hole} pockets.  

Here we show that the presence or absence of the sign reversal between the hole pockets can be inferred from doping variations of low-energy quasiparticle (QP) density of states (DOS).   We have measured the thermal conductivity and penetration depth,  both of which are very sensitive bulk probe for the low-energy QP excitations, down to very low temperature in the heavily hole doped regime of Ba$_{1-x}$K$_x$Fe$_2$As$_2$.  The results suggest that the sign change of SC gap between the hole pockets occurs very close to $x=1$, implying the importance of the inter-hole-pocket scattering for the electron pairing.   

Single crystals of Ba$_{1-x}$K$_x$Fe$_2$As$_2$ have been grown by the KAs flux method \cite{Kihou}.  For $x=1$, 0.93, 0.88 and 0.76, the $T_c$s are 3.7, 7.3, 11 and 22\,K, residual resistivity ratios $RRR$ are 1906, 175, 206 and 49, and the upper critical field $H_{c2}$($\parallel c$ axis) at $T=0$\,K are 1.8, 6.5, 17.3 and 28\,T, respectively.   For $x=0.88$ and 0.76, $H_{c2}(0)$  are estimated by assuming Werthamer-Helfand-Hohenberg relation \cite{WHH}.  Thermal conductivity is measured by the steady-state method and the penetration depth is determined by the tunnel diode oscillator technique \cite{Hashimoto102}. 

\begin{figure}[t]
\includegraphics[width=0.9\linewidth]{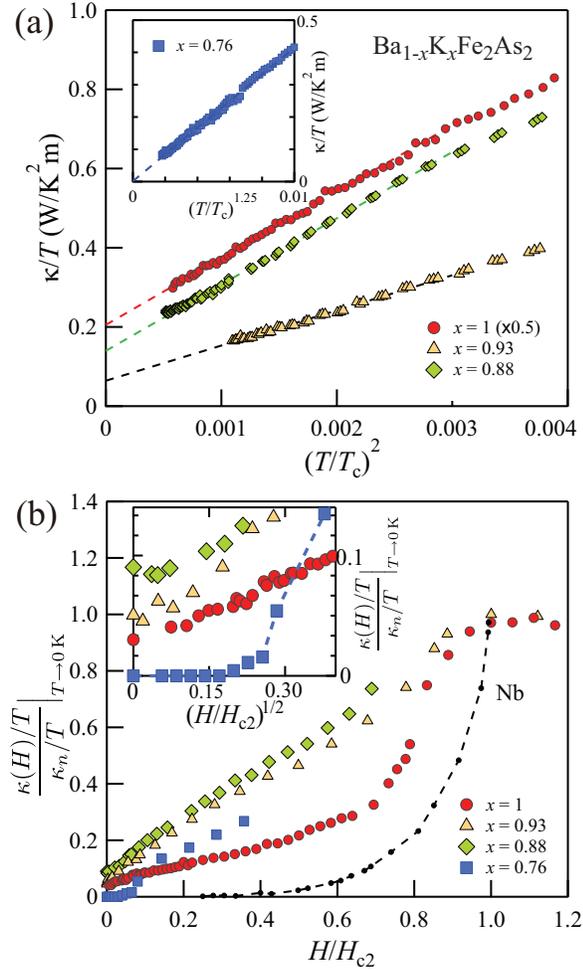}

\caption{(Color online). Low-temperature thermal conductivity in Ba$_{1-x}$K$_x$Fe$_2$As$_2$ single crystals. (a) $\kappa/T$ vs $(T/T_c)^2$ for $x=1.0$, 0.93, 0.88. Inset shows $\kappa/T$ plotted against $(T/T_c)^{1.25}$ for $x=0.76$. (b) Field dependence of $\kappa/T$ normalized to the normal-state values $\kappa_n/T$. Each point is determined from the extrapolation of the temperature dependence to $T\rightarrow 0$\,K at each field applied along the $c$ axis. For comparison, the data for Nb is also plotted \cite{Lowell70}. Inset is an expanded view at low fields. Dashed lines are the guide to the eyes. 
}
 \label{kappa}
\end{figure}

There is a fundamental difference in the low-temperature thermal transport between superconductors with isotropic and nodal gaps.  In the latter the QP heat conduction is entirely governed by the low-energy nodal excitations.
Figure\:\ref{kappa}(a) shows the zero-field thermal conductivity divided by temperature $\kappa/T$ well below $T_c$ plotted as a function of $(T/T_c)^2$ for  $x$=0.88, 0.93 and 1.  The total thermal conductivity is a sum of the electron and phonon contributions. The former is represented by $\kappa_e/T \approx N(0)v_F^2\tau_e$, where $N(0)$, $v_F$,  and $\tau_e$ are the QP DOS, Fermi velocity and QP scattering time, respectively.  At low temperatures, $\kappa/T$ is well fitted by $\kappa/T=\tilde\kappa_0 +bT^2$ , where $\tilde\kappa_0$ and $b$ are constants.  The finite residual value in $\kappa/T$ at $T \rightarrow 0$\,K, $\tilde\kappa_0$,  is clearly resolved.  This indicates the presence of the residual QP DOS, which can be attributed to be the presence of line nodes in the SC gap function \cite{Graf96,Mishra09}.  In sharp contrast, as depicted in the inset of of Fig.\:\ref{kappa}(a),  $\kappa/T$ for $x=0.76$ does not show $T^2$ dependence but is well fitted by $\kappa/T\propto T^{1.25}$ with no residual value, indicating a full SC gap.  

Another strong indication for the line node for $x=1$, 0.93 and 0.88 and the full SC gap for $x=0.76$ is provided by the field dependence of the thermal conductivity.   In magnetic field the thermal transport of superconductors with line node is dominated by contributions from delocalized QP states outside the vortex cores.  Then $N(0)$ increases in proportion to $\sqrt{H}$ (Volovik effect), which gives rise to a steep increase of $\kappa(H)$ at $H\ll H_{c2}$ \cite{Matsuda06}.  On the other hand, in fully gapped superconductors, all QPs are localized within the vortex cores and unable to transport heat.  Then $\kappa(H)$ exhibits an exponential behavior with a very small growth with $H$ at low field (see the data of Nb \cite{Lowell70} in Fig.\:\ref{kappa}(b)).   Figure\:\ref{kappa}(b) shows the field dependence of  $\kappa/T$ normalized by the normal-state value $\kappa_n/T$ at $T \rightarrow 0$\,K for {\boldmath $H$}$\parallel c$.     For $x=0.88$ and 0.76,  $\kappa_n$ is determined by the normal state resistivity extrapolated to $T\rightarrow 0$\,K assuming the Wiedemann-Franz law.  The field dependence of $\kappa/T$ for $x=0.76$ is fundamentally different from that for other systems: At low fields, $\kappa/T$ for $x=1.0$, 0.93, 0.88,  displays a convex field dependence, while concave one is observed for $x=0.76$.  This can be seen more clearly in the inset of Fig.\:\ref{kappa}(b), which shows a zoom of the low field region.   For $x=1.0$, 0.93 and 0.76,  $\kappa(H)/T (\propto N(H))$ increases in proportional to $\sqrt{H}$, indicating the presence of line node.   In contrast, $\kappa(H)/T$ for $x=0.76$ stays nearly zero  up to $\sqrt{H/H_{c2}}\sim 0.16$,  followed by a rapid increase at around $\sqrt{H/H_{c2}}\sim 0.3$, indicating the fully gapped superconductivity.

\begin{figure}[t]
\includegraphics[width=0.9\linewidth]{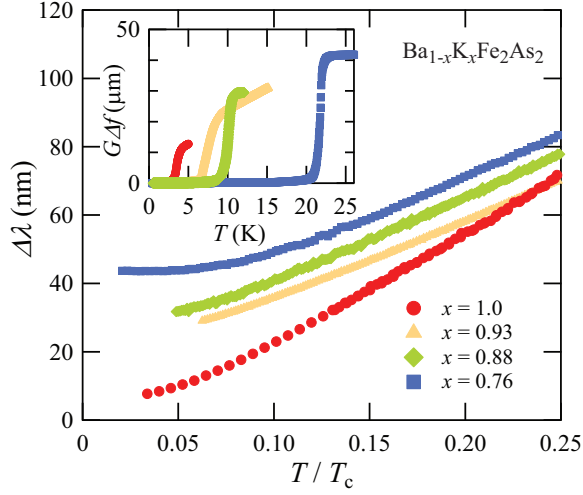}
\caption{(Color online). Change in the penetration depth as a function of $T/T_c$ in Ba$_{1-x}$K$_x$Fe$_2$As$_2$. Each curve is vertically shifted for clarity. Inset: the temperature dependence of frequency shift $\Delta f$ multiplied by the geometrical factor $G$ measured by the tunnel diode oscillator \cite{Hashimoto102} in a wide temperature range shows sharp superconducting transition. The onset of the Meissner signal almost coincides with the transition temperature $T_c$ determined by zero resistivity. In the superconducting state, the penetration depth can be deduced from $\Delta\lambda=G\Delta f$. 
} \label{lambda}
\end{figure}

A further test of the absence or presence of line node is provided by London penetration depth $\lambda$.  Figure\:\ref{lambda} depicts the temperature dependence of the relative change of $\lambda(T)$,  $\Delta \lambda(T) = \lambda(T)-\lambda(0)$, at low temperatures plotted against $T/T_c$.   The data for $x=0.76$ is completely flat within the experimental error of  $\sim 0.3$\,nm below $T/T_c < 0.05$, indicating negligible QP excitations, i.e. a fully gapped SC state.  The data for $x=1.0$, 0.93 and 0.88, on the other hand,  exhibits steeper temperature dependence of $\Delta \lambda (T)$.  When we use a power law fit  $\Delta \lambda (T) \propto T^{\alpha}$ to these data below $T/T_c \sim 0.25$, we obtain a small value of $\alpha \approx 1.3-1.5$.   These small powers $\alpha<1.5$  cannot be explained by a dirty nodeless state, and it is rather a strong indication of line nodes \cite{Hashimoto10}.  

Thus the $T$- and $H$-dependencies of the thermal conductivity and the penetration depth, all consistently indicate the SC gap with line nodes for $x=1.0$, 0.93 and 0.88 and full SC gap for $x$=0.76.   Compared to conventional superconductor Nb, $\kappa(H)/T$ for $x=0.76$ exhibits a rapid increase at much lower $H/H_{c2}$.  Moreover, a flat region of $\Delta \lambda(T)$  is achieved only below $T/T_c < 0.05$.  These indicate that the SC gap function for $x=0.76$ is strongly modulated and the minimum gap value is very small.  This implies that the line node appears slightly above $x=0.76$.  

\begin{figure}[t]
\includegraphics[width=\linewidth]{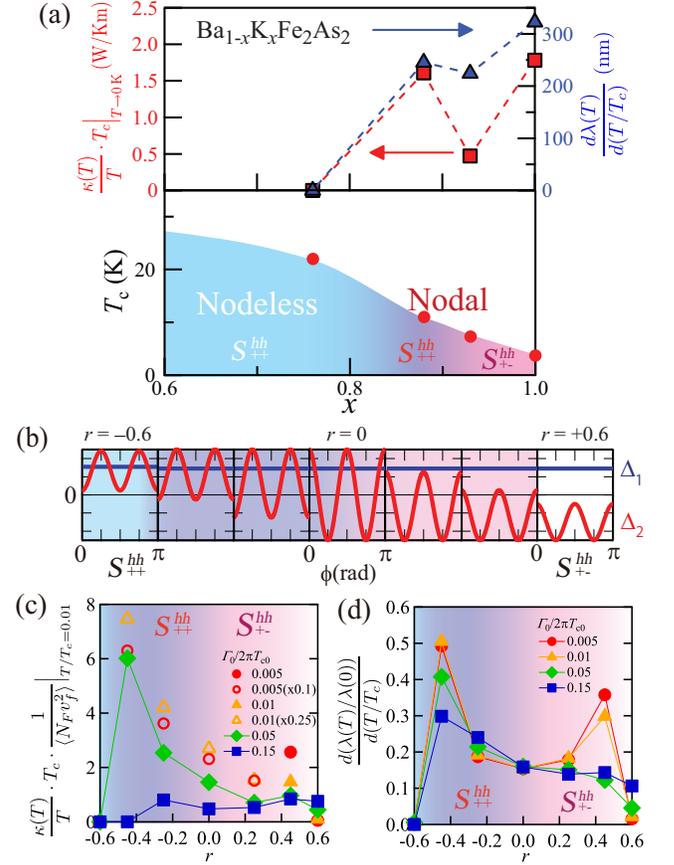}
\caption{(Color online). Doping evolution of low-energy QP excitations in Ba$_{1-x}$K$_x$Fe$_2$As$_2$. (a) Upper panel: $x$-dependence of $\left. \frac{\kappa(T)}{T}\cdot T_c \right|_{T\rightarrow 0\,\rm{K} }$(red squares, left axis) and $\frac{d\lambda(T)}{d(T/T_c)}$ (blue triangles, right axis) determined by the slope in a range of $0.1<T/T_c<0.2$ for $x\geq 0.88$ . Lower panel: $x$ dependence of $T_c$.  (b) We associate the above non-monotonic behavior with the gap evolution from predominantly $S^{hh}_{+-}$($r=0.6$) to predominantly $S^{hh}_{++}$($r=-0.6$) through intermediate regime with accidental nodes in $\Delta_2(\phi)$, as found in the laser ARPES measurements for $x=1$ \cite{Okazaki12}. (c)(d) Theoretically determined residual $\kappa(T)/T$ (c), and $d\lambda(T)/dT$ (d) for gap evolution shown in (b) with several values of impurity scattering $\Gamma_0$.  They show a characteristic double hump non-monotonic structure consistent with experimental findings. 
} \label{doping}
\end{figure}

We are now in  a position to discuss the issue of sign reversal of the SC gap between the hole pockets.  Figure\:\ref{doping}(a) depicts the doping evolution of $\left. \frac{\kappa}{T}\cdot T_c \right|_{T\rightarrow 0}\,K$ and $\frac{d \lambda}{d(T/T_c)}$.   Some intuition about their behavior can be obtained based on the clean limit results for $d$-wave \cite{Graf96} and anisotropic $s$-wave \cite{Mishra09} that look similar: 
\begin{equation}
(\kappa/T)\cdot  T_c|_{T\rightarrow 0\,{\rm K}}= aN_Fv_F^2\mu^{-1},
\qquad 
(d\lambda/dT) \cdot T_c \propto \mu^{-1}.
\end{equation}
Here $a$ is a constant of order unity and $N_F$ is the normal state DOS, and 
the transport time, $1/\mu$, is inverse `gap velocity' 
$\left. \mu=\frac{1}{\Delta}\frac{d|\tilde\Delta(\phi)|}{d\phi} \right|_{node}$
that describes opening of the SC gap at the node
(here $\Delta$ is the $T_c$-related gap scale, and $|\tilde \Delta(\phi)|$ is the impurity renormalized gap). 

As reported by the NMR Knight shift measurements \cite{Hirano12},  $N_F$ is nearly $x$-independent in the present $x$-range.  From this also follows that it is unlikely that $v_F$ strongly depends on $x$.  The nodal parameter $\mu$ is, therefore,  the main factor responsible for the $x$ dependence of  $\left. \frac{\kappa}{T}\cdot T_c \right|_{T\rightarrow 0\,{\rm K}}$ and $\frac{d\lambda}{d(T/T_c)}$.  However, to determine   this $\mu$ parameter, a comparison with theory is needed, as it is strongly affected by the impurity renormalized gap $|\tilde \Delta(\phi)|$ in case of accidental nodes \cite{Mishra09}.

The most remarkable feature in Fig.\:\ref{doping}(a) is that both $\left. \frac{\kappa}{T}\cdot T_c \right|_{T\rightarrow 0\,{\rm K}}$ and $\frac{d\lambda}{d(T/T_c)}$ exhibit a similar doping dependence.   As $x$ is decreased from 1, both of them first decrease, go through a minimum at $x\sim 0.93$, increase to a maximum at $x \sim 0.88$, and then decrease again. Both are expected to vanish at $x$ slightly above 0.76 where nodes disappear.  The observed non-monotonic $x$-dependence in a narrow doping range is directly connected with unusual evolution of the low-energy QP excitations.  We first point out that these results provide a strong support for the nodal $s$-wave symmetry for $x \geq 0.88$, rather than $d$-wave. 
This is because in symmetry-protected $d$-wave superconductors  $\mu$ and hence both $\left. \frac{\kappa}{T}\cdot T_c \right|_{T\rightarrow 0\,{\rm K}}$ and $\frac{d\lambda}{d(T/T_c)}$, are expected to be mostly independent of doping and impurity level.  On the other hand, gap with accidental nodes has nothing special about its structure and is expected to depend strongly on  both \cite{Hirschfeld11,Mishra09}. 

We stress that the observed non-monotonic doping evolution of  $\left. \frac{\kappa}{T}\cdot T_c \right|_{T\rightarrow 0\,{\rm K}}$ and $\frac{d\lambda}{d(T/T_c)}$ provides an important hint for the SC order parameter in the heavily hole-doped regime, and suggests a crossover from an $s$-wave state with sign reversed order parameter between the hole bands ($S^{hh}_{+-}$) to $s$-wave state without sign reversal ($S^{hh}_{++}$).  Here we consider the SC gap evolution shown in Fig.\:\ref{doping}(b), assuming a model with different gap functions, isotropic in band 1 and anisotropic in band 2 \cite{ARPES}: 
\begin{equation}
\begin{array}{l}
\Delta_1(\phi)=\Delta_{10}, \\
\Delta_2(\phi)= -\mbox{sgn}(r) \Delta_{20} \{|r| - (1-|r|)\cos4\phi\}.
\end{array}
\end{equation}
The positive $r$ represents an $S_{+-}^{hh}$ state with a sign change between the hole pockets and the negative $r$ represents a  $S_{++}^{hh}$ state with no sign change.
For $|r|<0.5$ accidental nodes appear in $\Delta_2(\phi)$, and a crossover from  $S_{+-}^{hh}$ to  $S_{++}^{hh}$ state occurs continuously at $r=0$. Theoretically we self-consistently compute the order parameter and, more importantly, impurity self-energies with the quasiclassical two-band approach using t-matrix, outlined in \cite{Mishra09,Das13}. We use unitary scattering limit 
and for the case of two $\Gamma$-centered hole bands we assume 
uniform scattering, $V_{interband}=V_{intraband}$.  Impurity self-energies in two bands are used to calculate the thermal conductivity $\kappa(T) = \kappa_1(T)+\kappa_2(T)$ 
using Keldysh formalism \cite{Graf96,Anton07ii}. 
The results of the theoretical model for various gap structures are presented in Figs.\:\ref{doping}(c) and (d) for several values of impurity scattering $\Gamma_0$. 

In the optimally doped sample ($x=0.45$) both theory and experiment point to 
the hole-electron $S^{he}_{+-}$ state with same-sign gap on the hole pockets, $S^{hh}_{++}$. 
There is no evidence for another nodal state appearing in the intermediate doping region between $x=0.45$ and 0.76 \cite{Malaeb12,Nakayama11}, and thus we consider that $x=0.76$ sample has an $S_{++}^{hh}$ state as well, but possibly with strongly anisotropic $\Delta_2(\phi)$.  The fact that the line node disappears slightly above $x=0.76$ suggests that the gap function of $x=0.76$ is close to $r=-0.6$ in our model.  On the other end, $x \approx 1$, the laser ARPES measurement in KFe$_2$As$_2$ \cite{Okazaki12}
points to gap structure similar to either $r=-0.4$ ($S_{++}^{hh}$) or 0.4 ($S_{+-}^{hh}$).
So the possible gap evolution with doping between $x=1$ and $x=0.76$ 
can either go (a) from $r=-0.4 \to -0.6$, or (b) from $r=0.4 \to -0.6$. 
In (a) it is a transition between very similar gap structures $S_{++}^{hh} \to S_{++}^{hh}$, where one expects at most a monotonic change in the nodal parameter $\mu$, and consequently monotonic low-$T$ observables.  In (b), on the other hand, a crossover from $S_{+-}^{hh}$ to $S_{++}^{hh}$ state occurs. If one considers the $x=1$ order parameter to correspond in our model to $r=0.45$, and  taking into account that the sample in $x=1$ case was much cleaner than at other dopings,  the theoretical non-monotonic doping dependence of  $\left. \frac{\kappa}{T}\cdot T_c \right|_{T\rightarrow 0\,{\rm K}}$ and $\frac{d\lambda}{d(T/T_c)}$ is consistent with experiment, and points to \emph{the sign reversal between the hole pockets} in KFe$_2$As$_2$.  

We also note that the calculation results indicate that the residual thermal conductivity only weakly depends on impurity concentration on the $S_{+-}^{hh}$ side, and very sensitive  to it on the $S_{++}^{hh}$ side, c.f. cases $r= \pm 0.45$. This, together with the reported results that a couple of $x=1$ samples with different residual resistivity show similar values of $\left. \frac{\kappa}{T}\cdot T_c \right|_{T\rightarrow 0\,{\rm K}}$ \cite{Reid12,Wang12}, gives a fully consistent picture that in  the $A_{1g}$ state, inferred from laser ARPES \cite{Okazaki12}, the $\Gamma$-centered hole bands have order parameter of predominantly opposite signs.

In iron-pnictides with both electron and hole pockets,  the electron-hole interaction is crucially important for the electron pairing.  According to the antiferromagnetic spin fluctuation senario, this interaction gives rise to the sign change between these pockets \cite{Hirschfeld11,Kontani10}.  Without the electron pockets, however, it seems that the hole-hole interaction steps up to the front stage, as revealed by the sign reversal between hole pockets in KFe$_2$As$_2$.  The present results imply that several competing interactions (electron-hole and hole-hole, or simply hole-hole in the absence of electron pockets), give rise to a unique and rich variety of SC gap structure of iron-pnictides, very different from that of all other superconductors. 

In summary,  from  the  thermal conductivity and penetration depth measurements of Ba$_{1-x}$K$_x$Fe$_2$As$_2$ in the heavily hole doped regime, we show that  the slope of the gap at the node exhibit a highly unusual  non-monotonic $x$-dependence.  These results imply that the nodal gap function of KFe$_2$As$_2$ is not symmetry-protected but has extended $s$-wave  structure and reverses sign between the hole pockets, suggesting  that the inter-hole-pocket scattering plays an important role in superconductivity.

{\it Note:} After we have finished this work we became aware of very recent ARPES results which show a change from nodal to nodeless state between dopings corresponding to samples with $T_c\sim 17$\,K and 22\,K \cite{Ota13}, consistent with our results. 

We thank A.\,V. Chubukov, P.\,J. Hirschfeld, H. Kontani, K. Okazaki, Y. Ota, and S. Shin for valuable discussions. This work has been supported by KAKENHI from JSPS. 
A.B.V. acknowledges support from NSF through grant DMR-0954342. 



\begin{thebibliography}{99}
\bibitem{ARPES} An almost-zero SC gap has been reported in the outer hole pocket in KFe$_2$As$_2$ at 2\,K \cite{Okazaki12}.   Such a small gap, however, is not observed in the penetration depth and $H$-dependnece of thermal conductivity measurements at very low temperatures, implying that this gap has no node and grows rapidly at low temperature or this band gives much smaller contribution than other bands to these quantities. We therefore simply assume two gap functions given by Eq.\,(2).
\end{thebibliography}
\end{document}